\documentclass[useAMS,usenatbib,usedcolumn,psfig,eps,graphics]{mn2e} 
\usepackage{epsfig, graphicx,natbib} 
 
\def\grtsim{\mathrel{\hbox{\rlap{\hbox{\lower2pt\hbox{$\sim$}}}\raise2pt\hbox{$>$}}}} 
\def\lesssim{\mathrel{\hbox{\rlap{\hbox{\lower2pt\hbox{$\sim$}}}\raise2pt\hbox{$<$}}}}

\def\degree{\nobreak\ifmmode{^\circ}\else{$^\circ$}\fi}

\newcommand{\mc}{\multicolumn}

\newcommand{\aap}{A\&A}

\newcommand{\aj}{AJ} 
\newcommand{\apj}{ApJ} 
\newcommand{\apjl}{ApJL} 
\newcommand{\apjs}{ApJS} 
 
\newcommand{\araa}{ARA\&A} 
\newcommand{\mnras}{MNRAS} 
\newcommand{\nat}{Nat}

\newcommand{\pasp}{PASP}

\begin{document} 
\topmargin -0.5in 
 
\title[A population of high-$z$ type-2 quasars-II ]{A population of high-redshift type-2 quasars-II. Radio Properties} \author[A. Mart\'\i 
nez-Sansigre et al. ]{Alejo Mart\'\i nez-Sansigre$^{1}$\thanks{E-mail: 
a.martinez-sansigre1@physics.oxford.ac.uk (AMS)},  Steve Rawlings$^{1}$,  Timothy  Garn$^{2}$, David A. Green$^{2}$, \and 
Paul Alexander$^{2}$, Hans-Rainer Kl\"ockner$^{1}$, Julia M. Riley$^{2}$ 
\\  
\footnotesize\\  
$^{1}$Astrophysics, Department of Physics,University of Oxford, Keble Road, Oxford OX1 3RH, UK\\ 
$^{2}$Astrophysics Group, Cavendish Laboratory, Madingley Road, Cambridge CB3 0HE\\ 
}

\date{\today}

\pagerange{\pageref{firstpage}--\pageref{lastpage}} \pubyear{} 
 
\maketitle 
 
\label{firstpage}

\begin{abstract}  {We present multi-frequency radio observations of a sample of $z\sim2$ obscured (type-2) quasars in the Spitzer extragalactic First Look Survey area. We combine the public data at 1.4 GHz, used in the selection of these sources, with new observations at 610 MHz (GMRT) and at 4.9 GHz (VLA). We find the sample includes sources with steep, flat  and gigahertz-peaked spectra. There are no strong correlations between the presence or absence of emission lines in the optical spectra and the radio spectral properties of the sample. However, there are no secure flat-spectrum type-2 quasars with narrow emission lines which would be problematic for unified schemes.  Most of the population have straight radio spectra with spectral index $\alpha \sim 1$ as is expected for developed, potentially FRI-like, jets in which continous injection of relativistic electrons is accompanied by inverse-Compton losses against the cosmic microwave background.    }

\end{abstract} 
 
 
\section{Introduction} 
 
The `zoo' of active galactic nuclei (AGN) includes many different 
objects with signs of non-stellar activity, powered by a supermassive 
black hole at the centre of a host galaxy 
\citep{1984ARA&A..22..471R}. The different observed properties are 
usually explained via variations in the rate of accretion of this 
central black hole, the black hole mass, and the orientation of the 
axes of symmetry with respect to the observer 
\citep[e.g.][]{1993ARA&A..31..473A}.  Obscured (type-2) quasars are 
intrinsically the same as unobscured (type-1) quasars, except for the 
orientation, where the dusty torus around the central black hole 
blocks the line of sight to the bright accretion disk and the 
broad-line region. Only the narrow-line region is then visible to the 
observer. Radio-loud type-2 quasars, in the form of narrow emission 
line radio galaxies, have been known for a long time. For the more 
radio-luminous galaxies \citep[the FRII class of ][] 
{1974MNRAS.167P..31F}, convincing evidence exists for their obscured 
quasar nuclei \citep[e.g.][] {2005A&A...442L..39H,Ogle:2006cu}.  However, not 
all FRII radio-galaxies are powerful accreters \citep{Ogle:2006cu}, 
and amongst the less radio-luminous FRI class few quasars were found 
\citep [see ] [ for an exception] {2001ApJ...562L...5B}, although 
recent radio imaging of quasars has challenged this (Heywood, Blundell 
\& Rawlings, MNRAS, submitted). Amongst the radio-quiet population, the 
type-2 quasars had remained elusive until recently.

The search for high-redshift type-2 quasars has been carried out 
mainly in the X-rays \citep [see][for a review] {2005ARA&A..43..827B} 
and, more recently, in the mid-infrared \citep 
[e.g.][]{2004ApJS..154..166L,2005Natur.436..666M}.  The sample of 
\citet {2005Natur.436..666M,2006MNRAS.370.1479M} consists of $z\sim2$ 
radio-intermediate type-2 quasars: quasars with radio luminosities 
below the FRI/FRII break, but above that would be expected from 
star-formation alone. These were selected in the Spitzer extragalactic 
First Look Survey (FLS) by demanding objects to have a 24-$\mu$m flux 
density $S_{24}>300~\mu$Jy, a 3.6-$\mu$m flux density 
$S_{3.6}\leq45~\mu$Jy, and a 1.4 GHz flux density in the range 
$350~\mu$Jy $\leq S_{1.4} \leq$ 2 mJy.  \citet {2006MNRAS.370.1479M} 
found secure optical redshifts (from narrow lines) for $\sim50$\% of 
their sample with the remaining spectra typically completely blank: 
they discuss the possibility of their blank optical spectra being due 
to a dusty host galaxy obscuring all the central activity, even the 
narrow emission lines. Similar type-2s with blank optical spectra are 
found in the samples of \citet {2004ApJS..154..166L} and \citet 
{2005A&A...440L...5L}.

\begin{table*} 
\begin{center} 
\begin{tabular}{lllcrrrrrr} 
\hline 
\hline 
Name & RA & Dec &   $S_{4.9 \rm GHz}$ &  $S_{1.4 \rm GHz}$ & $S_{610 \rm MHz}$ & $\alpha^{4.9}_{1.4}$ & $\alpha^{1.4}_{610}$ & $z_{\rm spec}$ & Spec \\ 
 
 & (J2000)  &  &  \mc{1}{c}{/ $\mu $Jy} & \mc{1}{c}{/ $\mu $Jy} & \mc{1}{c}{/ $\mu $Jy} & &  \\ 
\hline 
AMS01 & 17 13 11.17 & +59 55 51.5  & 164 $\pm$ 52 & 490  $\pm$ 31 & 1167 $\pm$ 165 & 0.87$^{+0.35}_{-0.27}$ & 1.04$^{+0.13}_{-0.13}$ & - &1 \\  
AMS02 & 17 13 15.88 & +60 02 34.2  & 408 $\pm$ 52  & 1184 $\pm$ 55 & 2985 $\pm$ 152 & 0.85$^{+0.15}_{-0.13}$ & 1.11$^{+0.08}_{-0.08}$ & - &1 \\  
AMS03 & 17 13 40.19 & +59 27 45.8  &  365 $\pm$ 53  & 1986 $\pm$ 87 & 5180 $\pm$ 191 & 1.35$^{+0.16}_{-0.14}$ & 1.15$^{+0.07}_{-0.07}$ & 2.698 &2 \\  
AMS04A$^{\dag}$ & 17 13 40.62 & +59 49 17.1  &  135 $\pm$ 53  & 536  $\pm$ 72 & 1092 $\pm$ 77 & 1.10$^{+0.50}_{-0.38}$ & 0.86$^{+0.26}_{-0.24}$ & 1.782 &2 \\  
AMS04B$^{\ddag}$ & 17 13 40.62 & +59 49 17.1  &  232 $\pm$ 106  & 867  $\pm$ 101 & 1519 $\pm$ 122 & 1.05$^{+0.58}_{-0.40}$ & 0.68$^{+0.24}_{-0.23}$ & 1.782 &2 \\  
AMS05 & 17 13 42.77 & +59 39 20.2  &  443 $\pm$ 51  & 1038 $\pm$ 49 & 1214 $\pm$ 164 & 0.68$^{+0.14}_{-0.13}$ & 0.19$^{+0.13}_{-0.13}$ & 2.017 &2 \\  
AMS06 & 17 13 43.91 & +59 57 14.6  & 191$\pm$53   & 444  $\pm$ 31 & 1115 $\pm$ 153 &0.67$^{+0.31}_{-0.25}$ & 1.11$^{+0.14}_{-0.14}$ & 1.76  &1,3 \\  
AMS07 & 17 14 02.25 & +59 48 28.8  &  239 $\pm$ 51  & 354  $\pm$ 27 & 578  $\pm$ 151 & 0.31$^{+0.25}_{-0.22}$ & 0.59$^{+0.20}_{-0.20}$ & - &1 \\  
AMS08 & 17 14 29.67 & +59 32 33.5  & $<$718   & 655  $\pm$ 36 & 889  $\pm$ 151 &$>$-0.07 & 0.37$^{+0.21}_{-0.22}$ & 1.979 &2 \\  
AMS09 & 17 14 34.87 & +58 56 46.4  &  174 $\pm$ 51  & 426  $\pm$ 29 & 1118 $\pm$ 190 & 0.71$^{+0.33}_{-0.26}$ & 1.16$^{+0.18}_{-0.18}$ & - &1 \\  
AMS10 & 17 16 20.08 & +59 40 26.5  & 501$\pm$ 36  & 1645 $\pm$ 73 & 4590 $\pm$ ~93 & 1.04$^{+0.13}_{-0.12}$ & 1.24$^{+0.74}_{-0.07}$ & - &1 \\  
AMS11 & 17 18 21.33 & +59 40 27.1 & $<$104  & 356  $\pm$ 29 & 628  $\pm$ ~89 &$>$0.98 & 0.68$^{+0.21}_{-0.21}$ & - &1 \\ 
AMS12 & 17 18 22.65 & +59 01 54.3  &  305 $\pm$ 83  & 946  $\pm$ 45 & 2440 $\pm$ 194 & 0.90$^{+0.19}_{-0.17}$ & 1.14$^{+0.10}_{-0.94}$ & 2.767 &2 \\  
AMS13 & 17 18 44.40 & +59 20 00.8  &  495 $\pm$ 80  & 1888 $\pm$ 83 & 4559 $\pm$ 135 & 1.07$^{+0.18}_{-0.16}$ & 1.06$^{+0.07}_{-0.72}$ & 1.974 &2,3 \\  
AMS14 & 17 18 45.47 & +58 51 22.5  &  158 $\pm$ 52  & 469  $\pm$ 29 & 1099 $\pm$ 196 & 0.87$^{+0.37}_{-0.28}$ & 1.03$^{+0.14}_{-0.14}$ & 1.794 &2 \\  
AMS15 & 17 18 56.93 & +59 03 25.0  & 326 $\pm$ 53  & 440  $\pm$ 29 & 570  $\pm$ 174 & 0.24$^{+0.19}_{-0.17}$ & 0.31$^{+0.23}_{-0.25}$ &  &4 \\  
AMS16 & 17 19 42.07 & +58 47 08.9  &  $<$108  & 390  $\pm$ 60 & $<$170   & $>$1.02 &$<$-1.00 & 4.169 &2 \\ 
AMS17 & 17 20 45.17 & +58 52 21.3  &  265 $\pm$ 53  & 615  $\pm$ 34 & 1354 $\pm$ 226 & 0.67$^{+0.22}_{-0.19}$ & 0.95$^{+0.18}_{-0.19}$ & 3.137 &2 \\  
AMS18 & 17 20 46.32 & +60 02 29.6  &  212 $\pm$ 53  & 390  $\pm$ 29 & 502  $\pm$ 113 & 0.49$^{+0.29}_{-0.24}$ & 0.30$^{+0.25}_{-0.26}$ & 1.017 &2 \\  
AMS19 &  17 20 48.00 & +59 43 20.7  &  506 $\pm$ 52  & 822  $\pm$ 41 & 958  $\pm$ 152 & 0.50$^{+0.18}_{-0.16}$ & 0.18$^{+0.17}_{-0.17}$ & 2.25 &1,3 \\ 
AMS20 & 17 20 59.10 & +59 17 50.5  & 1391 $\pm$53  & 1268 $\pm$ 58 & 674  $\pm$ 187 & -0.07$^{+0.07}_{-0.07}$ &-0.76$^{+0.32}_{-0.40}$ & - &1 \\  
AMS21 & 17 21 20.09 & +59 03 48.6  & 145 $\pm$ 53  & 449  $\pm$ 29 & 1019 $\pm$ 169 & 0.91$^{+0.42}_{-0.30}$ & 0.99$^{+0.16}_{-0.16}$ & - &1 \\  
\hline 
\hline 
\end{tabular} 
\caption{ \noindent Positions, redshifts and radio data for the entire 
sample. The integrated fluxes with errors are from the AIPS tasks 
IMFIT, SAD and JMFIT (for 4.9 GHz, 1.4 GHz and 610 MHz 
respectively). The last column states the spectroscopic properties: 1 
blank in optical spectrum, 2 shows narrow lines ($\lesssim 2000$ km 
s$^{-1}$) in the optical spectrum, 3 has a redshift determined from 
Spitzer-IRS, 4 has not been observed by optical spectroscopy. The 
optical redshifts obtained using the William Herschel Telescope-ISIS 
instrument are from \citet {2006MNRAS.370.1479M}, while the 
Spitzer-IRS redshifts are from \citet {2005ApJ...628..604Y} or \citet 
{2006ApJ...638..613W}. $^{\dag}$The properties of AMS04 without the flux density of the  
 source at 17 13 41.20 +59 49 24.0 which might be an extended component of AMS04. $^{\ddag}$The properties of AMS04 including the flux density of the possible extended component. } 
\label{tab:table} 
\end{center} 
\end{table*}

Radio observations can be used to test whether the type-2 quasars are 
obscured by the torus. The first expectation is that radio jets, if 
present, are at large angles to the line-of-sight and steep 
 radio spectral indices are expected. If the jets in 
some of the type-2 quasars were recently triggered, then GHz-peaked 
radio spectra, characteristic of young jets \citep 
{1998PASP..110..493O} might be observed. Classic type-2 quasars 
(presumably those showing narrow-lines) are not expected to have the 
jet close to the observer's line of sight and exhibit the flat radio 
spectra seen in some type-1 quasars. Therefore, the suggestion by 
Miller, Rawlings \& Saunders (1993) and Falcke, Malkan \& Biermann 
(1995) that the radio-intermediate quasar population is comprised of 
Doppler-boosted flat-spectrum radio-quiet quasars would pose serious 
difficulties for the narrow-line type-2 quasars in this 
sample. However, if some quasars are obscured by dust independent of 
the obscuration, as suggested by \citet {2006MNRAS.370.1479M}, then 
the jets could be close to the line of sight, and flat radio spectra 
might be observed. This predicts that a small fraction of the sources 
with no  narrow lines should have flat radio spectra, although 
this fraction will be boosted by a radio selection criterion. 
 
\vspace{-0.1cm} 
In this letter we report the results of observing the entire Mart\'\i 
nez-Sansigre et al. sample in the FLS at 610~MHz and 4.9 GHz, combined 
with public data at 1.4 GHz, and the conclusions on orientation or age 
that can be inferred.  Throughout this paper we adopt a $\Lambda$CDM 
cosmology with the following parameters: $h = H_{0} / (100 ~ \rm km ~ 
s^{-1} ~ Mpc^{-1}) = 0.7$; $\Omega_{\rm m} = 0.3$; $\Omega_{\Lambda} = 
0.7$.

\section{Observations and data reduction} 
 
\subsection{1.4 GHz dataset} 
 
The 1.4~GHz data, taken from \citet {2003AJ....125.2411C}, were part 
of the dataset used to select the sample in the first place \citep 
{2005Natur.436..666M}. The 1.4~GHz survey of the FLS was carried out 
using the Very Large Array (VLA) in B-configuration, with a 
synthesised beam size of $\approx$5$\times$3 arcsec$^{2}$ and a root 
mean square (rms) noise of $\sim$23 $\mu$Jy.  The type-2 quasars were 
selected in the flux density range $350~\mu$Jy $\leq S_{1.4} \leq$ 2 
mJy. At 1.4 GHz, all the sources are point-like, except for AMS16, 
which has a slightly extended radio structure and AMS04 which has an 
adjacent radio source North-East (NE) which could plausibly be a 
jet. Several of the sources were also observed in the WSRT survey of 
\citet {2004A&A...424..371M} and comparison of the flux densities 
suggested no hints of extended radio structure \citep [other than 
AMS04, see] [] {2006MNRAS.370.1479M}.

\subsection{4.9 GHz dataset}

The 4.9~GHz data were also obtained using the VLA, but this time 
through pointed observations of every object in the sample, except 
AMS08 which was observed in the field of view of AMS03. Making use of 
dynamical scheduling time, the data presented here were effectively 
taken in C configuration. The observations were scheduled to optimise 
the coverage of the {\it uv} plane, the phase calibrators were visited 
every 9-12 minutes and 3C48 or 3C286 were used as amplitude 
calibrators. Overall, all sources were observed for almost exactly the 
same length of time, about 17 minutes each (except AMS10 which was 
observed for twice as long). With the VLA typically having $\sim$ 25 
working antennae each run, the expected thermal noise was $\sim$45 
$\mu$Jy ($\sim 32~\mu$Jy for AMS10).

The reduction of the 4.9~GHz data was peformed using the AIPS 
package. The sources themselves are too faint for self-calibration but 
some of the pointings include sources bright enough ($\sim$10 mJy) to 
allow tests for phase stability: the phases varied by 
$\lesssim$10\degree~in the 9-12 minute intervals between visits to the 
phase calibrator so, even where possible, self-calibration would have 
little effect on the final image. The beam size was 
$\approx$4$\times$3 arcsec$^{2}$, well matched to the 1.4~GHz beam, 
and the rms noise in the central region of all the images was found to 
be 50-54 $\mu$Jy (37 $\mu$Jy for AMS10), close to the expected thermal 
noise.  A primary beam correction was applied in the case of AMS08, 
since this source is at the edge of the image of AMS03. At 4.9~GHz, 
all detected sources are unresolved and point-like, except AMS04 which 
is slightly extended ($\approx6\times3$ arcec$^{2}$) in the NE to SW 
direction.

\subsection{610 MHz dataset}

The FLS region was observed in 2004 March with 7 pointings of the Giant Meterwave 
Radio Telescope (GMRT). The total exposure time per object amounted to 
$\approx 3.5$ hours. The flux 
density scale of the observations were tied to observations of 3C48 
and 3C286 (with assumed flux densities at 610~MHz of 29.4 and 21.1~Jy 
respectively) and the nearby source 1634$+$627 was used to monitor any 
phase and amplitude variations of the telescope. The data were 
calibrated and the images synthesised using standard techniques with 
the AIPS software. In this case self-calibration was possible on 
bright sources in the field of view. The rms noise on these images 
varied between 28 and 32 $\mu$Jy~beam$^{-1}$ before primary beam 
correction, although dynamic range limitations means that the noise 
close to bright sources is larger. Comparison of the flux densities of 
sources in the overlap regions between pointings, after correction for 
the nominal primary beam of the GMRT, revealed systematic 
inconsistencies between adjacent pointings. The differences could be 
explained by a consistent offset for the primary beam of the GMRT 
compared with the phase centre of each observation, of about 3 
arcmin. Consequently, an offset primary beam correction was made to 
each of the pointings, before determining flux densities for the 
sources.

At 610 MHz all sources are point-like, except AMS04. In this case the adjacent 
radio source is detected, and the gaussian fit to the combined object 
results in a 10.3 $\times$ 5.1 arcsec$^{2}$ ellipse, at a position angle of 40 
degrees (North through East).

\section{The spectral indices of the sample}\label{sec:indices}

The spectra of our sample are described here using two power laws, 
with flux density $S_{\nu} \propto \nu^{-\alpha}$. The spectral index 
between 1.4 and 4.9 GHz ($\alpha^{4.9}_{1.4}$) and between 610 MHz and 
1.4 GHz ($\alpha^{1.4}_{610}$) are given in Table~\ref{tab:table}. The 
flux densities used are all  integrated quantities from gaussian-fitting 
procedures.  Since the beam sizes at the three frequencies are well 
matched, and the sources are generally point-like, the spectral 
indices measured here are not strongly affected by problems such as 
confusion, or resolving out some of the flux density.

\begin{figure} 
\begin{center} 
\psfig{file=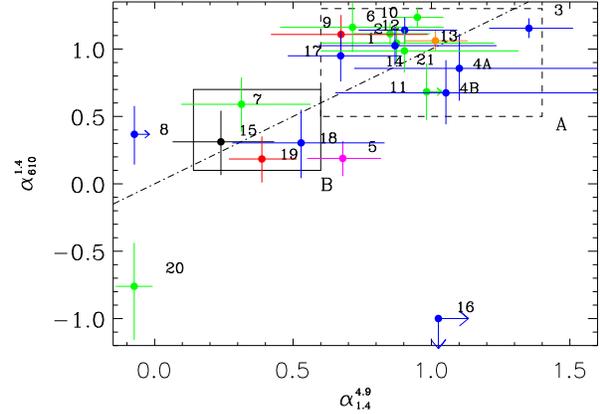,width=8cm,angle=0}  
\caption{\noindent Spectral index between 610~MHz and 1.4~GHz, 
 vs spectral index between 1.4 and 4.9 
GHz. The numbers identify each source: blue 
indicates sources with narrow-emission lines in their optical spectra; 
green, sources with blank optical spectra; the black point 
(AMS15) has not been observed with optical spectroscopy; red, the 
two objects (AMS06 and AMS19) with blank optical spectra, but 
spectroscopic redshifts from Spitzer-IRS; yellow shows AMS13, 
which has narrow lines in the optical, as well as a redshift from 
Spitzer-IRS; pink,   AMS05, with narrow lines in the 
optical, but no redshift obtained from Spitzer-IRS. The line 
$\alpha^{1.4}_{610}~=~\alpha^{4.9}_{1.4}$ and Boxes A and B have been drawn to aid discussion of Section~\ref{sec:indices} } 
\label{fig:indices} 
\end{center} 
\end{figure}

\begin{figure}
\begin{center} 
\psfig{file=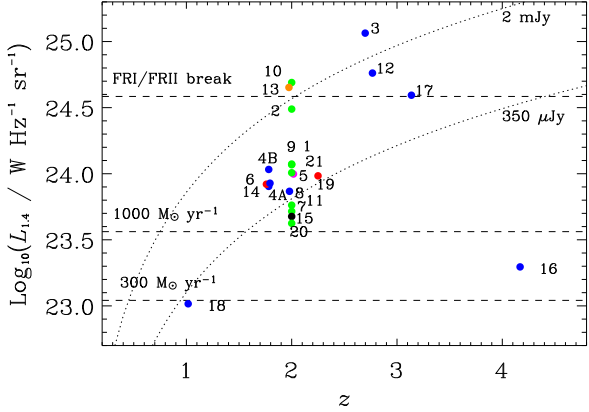,width=8cm,angle=0}  
\caption{\noindent The luminosity at 1.4~GHz plotted against 
redshift. The colour coding is the same as for 
Fig.~\ref{fig:indices}. The dotted lines indicate the radio criteria 
used in the selection of the sample and the dashed lines the FRI/FRII 
break and the radio luminosity due to star-formation rates of 1000 and 
300 M$_{\odot}$ yr$^{-1}$ in stars more massive than 5 M$_{\odot}$.} 
\label{fig:pz_plane} 
\end{center} 
\end{figure}

Table 1 summarises the data and Fig.~\ref{fig:indices} shows 
 $\alpha^{1.4}_{610}$ plotted against $\alpha^{4.9}_{1.4}$.  The 
 colours indicate the optical and mid-infrared spectroscopic 
 properties of the sources.  Fig.~\ref{fig:pz_plane} shows the 
 luminosity at 1.4~GHz as a function of redshift for the 
 sample. Objects with no spectroscopic redshift are placed at 
 $z=2$. The 1.4~GHz luminosities of the sources were calculated using 
 the spectral index $\alpha^{1.4}_{610}$ to convert from 
 observed-frame to rest-frame 1.4~GHz.  The dashed lines show the 
 luminosity of the FRI/FRII break \citep [$L_{178}=2\times10^{25}$ W 
 Hz$^{-1}$ sr$^{-1}$,][]{1974MNRAS.167P..31F}, converted from 178~MHz 
 to 1.4~GHz assuming $\alpha = 0.8$. Also plotted are the luminosities 
 expected for pure starbursts \citep [][also as dashed 
 lines]{1992ARA&A..30..575C}, and the two 1.4~GHz flux density cuts 
 used in our selection (dotted lines). For all these, we also assume 
 $\alpha = 0.8$. The difference in spectral indices between the 
 measured ones and the value assumed for the flux density cuts 
 explains why some sources fall outside the expected selection locus.

\subsection{Steep-spectrum sources} 
 
Looking at Fig.~\ref{fig:indices}, the majority of the sources have 
steep ($\alpha \sim 1$) indices in both intervals and the two indices 
have similar values. Considering objects in Box A, $\langle 
\alpha^{4.9}_{1.4}\rangle =0.91\pm0.18$ and $\langle 
\alpha^{1.4}_{610}\rangle =1.0\pm0.14$. This group includes objects 
with optical narrow lines and blank optical spectra. To test for any trends, we assume the hypothesis 
that all 13 sources in box A are realisations of the same underlying 
binomial distribution and have an $\sim$50\% chance of having 
narrow-emission lines; 6 out of 13 objects in Box A have emission 
lines which is obviously consistent with our hypothesis. Dividing the 
sources in two by splitting the Box by the median 
$\alpha^{4.9}_{1.4}$, $\alpha^{1.4}_{610}$ or by a diagonal median, we 
can test for trends, this time assuming the 6 sources in each half are 
drawn from this underlying distribution. In this case, deviations from 
90\% confidence would require one or zero sources of a given type, but 
each division always yields at least 2 sources of one or the other 
type. We can conclude that there is no trend or correlation between 
spectral index and optical properties in the sources in Box A.

In Fig.~\ref{fig:pz_plane}, we see several of these steep-spectrum 
sources (AMS03, AMS12, AMS13, AM17 and AMS10 if it is at $z \sim 2$) 
have radio luminosities around the FRI/FRII break. Since our objects are 
selected to be type-2 quasars, we have possibly found some highly 
accreting FRI radio sources, which although a small fraction of the 
low-$z$ FRI population, are not unknown \citep [see][and Heywood, Blundell \& Rawlings, MNRAS, submitted] {2001ApJ...562L...5B}.  In the case of AMS04, we 
tabulate and plot two sets of values, derived with (A) and without (B) 
the adjacent radio source at 17 13 41.20 +59 49 24.0 which could 
plausibly be a $\sim$10 arcsec jet associated with the type-2 
quasar. Including the jet lowers both indices, particularly 
$\alpha^{1.4}_{610}$, but the source is still clearly steep in both 
ranges.

Since the 1.4 GHz luminosities are similar to those of FRIs, we 
compare the spectral indices of our sample, to those compiled for the 
(178-MHz selected) 3C sample by \citet {1980MNRAS.190..903L}. At $z = 
2$, observed 610 MHz and 1.4 GHz correspond to rest-frame 1.8 and 4.2 
GHz respectively. So the spectral indices between 750 MHz and 5 GHz 
presented by \citet {1980MNRAS.190..903L}, for sources at $z \lesssim 
0.03$ include the range corresponding to our $\alpha^{1.4}_{610}$. 
They find typical spectral indices between 0.6 and 1.0 for FRI sources 
with values of $L_{1.4}$ between $10^{23.5}$ and $10^{25}$ W Hz$^{-1}$ 
sr$^{-1}$. The values of $\alpha^{1.4}_{610}$ presented here are 
clustered towards the high end of the Laing \& Peacock range. This is 
the reverse of what one expects from the most obvious selection effect, since the 
3C sources are selected at lower frequency, one expects them to have 
steeper indices than a 1.4 GHz-selected sample.  This could be 
explained by a combination of effects: e.g. the single-power-law 
spectra of Laing \& Peacock goes down to 750 MHz and spectral indices 
typically steepen at higher frequencies; the radio-source environments 
at low redshift are systematically different from those at high 
redshift; and the cosmic microwave background (CMB) energy density is 
significantly higher at $z\sim2$. 
 
Steep straight spectra with $\alpha\sim1$ are characteristic of the 
continuous injection of relativistic electrons accompanied by 
synchrotron \& inverse-Compton losses \citep {1962SvA.....6..317K}. We 
can speculate that the electrons are accelerated in the knots of 
FRI-like sources and have $\alpha\sim1$ because each has a 
(rest-frame) spectral `break frequency', $\nu_{\rm break}$ (where 
radiative losses steepen the negative slope of the number density of 
particles versus energy, and hence steepen the emitted spectrum) which 
has been driven below $\sim1$ GHz by the high value of the $z\sim2$ 
CMB energy density. The relevant equation \citep[Eqn. 3.10 
of][]{1991bja..book..100L} is 
 
\noindent\begin{equation} 
  \nu_{\rm break} = (9/4)\left( 1.12\times10^{3} B_{\rm source}\over{[B_{\rm source}^{2} + B_{\rm CMB}^{2}]^{2}t_{source}^{2}} \right) \rm nT^{3} \rm Myr^{2} 
\end{equation} 
 
\noindent where $B_{\rm CMB} = 0.318(1+z)^2$ nT, and $B_{\rm source}$ and 
$t_{source}$ are the characteristic magnetic field and age of the 
source respectively. At $z\sim2$ $B_{\rm CMB}\sim3$ nT. From minimum 
energy arguments, assuming $L_{1.4}=10^{24}$ W Hz$^{-1}$ sr$^{-1}$ and 
a characteristic size of 30 kpc, then $B_{\rm source}\sim1-2$ nT 
depending on the shape of the emitting region. Therefore a 10 Myr-old 
source at $z=2$ will have $\nu_{\rm break}\sim300$ MHz due to the high 
value of the CMB energy density.

\subsection{Flat-spectrum sources} 
 
Another visible group in Fig.~\ref{fig:indices} is Box B  which has $\alpha 
\lesssim 0.5$ in both radio spectral intervals. This includes the 
sources AMS15, AMS18 and AMS19, more arguably AMS07, and maybe AMS08. 
Flat ($\alpha \lesssim 0.5$) spectral indices in both frequency ranges 
suggest that the jets are close to the observer's line of sight with 
Doppler boosting of compact jet knots leading to a flat spectrum. In a 
unified scheme, such objects are not expected to be obscured by the 
torus, so any dust is expected to be on a larger scale, presumably in 
the host galaxy. This picture is consistent for AMS07 and AMS19 which 
have no narrow emission lines in their optical spectra, and could 
presumably be being obscured by dust in the host galaxy; AMS15 has not 
been observed spectroscopically. In Fig.~\ref{fig:pz_plane}, AMS07, 
AMS15 and AMS19 are seen to have 1.4~GHz luminosities consistent with 
radio-intermediate type-2 quasars: too high for star-formation only 
(unless they are hyper-luminous infrared galaxies) but below the 
FRI/FRII break.

AMS18 has a 1.4~GHz luminosity 
consistent with a pure starburst with a massive-star-formation rate of only $\sim 300$ M$_{\odot}$ yr$^{-1}$. Such an object could plausibly make it through our 
selection criteria and indeed AMS18 has an atypical optical spectrum showing only an  [O II] 3727~\AA~line and 
other low-ionization lines like C II] (2326~\AA) and Mg II 
(2798~\AA). No high-ionization lines are securely detected \citep [see 
Figure 6 of][]{2006MNRAS.370.1479M} so AMS18 could  well be a 
pure starburst. If the radio emission is indeed due to the starburst, 
then the flat spectral indices can be explained by 
a bremsstrahlung contribution.  
 
AMS08 has a flat spectrum between 1.4~GHz and 610~MHz 
($\alpha^{1.4}_{610} = 0.37$) but the limit obtained for $S_{4.9 \rm 
GHz}$ is of little use in constraining $\alpha^{4.9}_{1.4}$. Since it 
has narrow lines in the optical, a flat value for $\alpha^{4.9}_{1.4}$ 
would suggest a beamed jet and would not sit easily with the 
obscuration by the torus. However, a steep value ($\grtsim 0.7$) could 
be explained as a GHz-peaked source, in the same was as for 
AMS05 (see Section~\ref{sub:giga}).

\subsection {GHz-peaked sources}\label{sub:giga} 
 
Finally, there are three more sources with different spectral 
properties from those in Boxes A and B: AMS05; AMS16; and AMS20. These 
spectral properties can be explained as being due to young, compact 
jets, with synchrotron self-absorption leading to GHz-peaked spectra 
\cite [GPS; see][for a review]{1998PASP..110..493O}. The GPS sources 
reviewed by O'Dea have radio luminosities above the FRI/FRII break 
$L_{1.4}\geq4\times10^{25}$ W Hz$^{-1}$ sr$^{-1}$, so once again, the 
sources found here are radio-intermediate quasars as expected 
from the flux density cuts (see Figure~\ref{fig:pz_plane}). AMS20 has 
$\alpha^{1.4}_{610} = -0.76$ and $\alpha^{4.9}_{1.4} = -0.07$ 
suggesting the spectrum peaks somewhere above observed 4.9~GHz. We do 
not have a spectroscopic redshift for this object since the optical 
spectrum shows faint red continuum only but it is very unlikely to be 
at $z < 1$, so the peak is likely to be at a rest-frame frequency of 
9.8~GHz or higher. AMS16 (at $z = 4.169$) has a very negative 
$\alpha^{1.4}_{610}$ and a very positive $\alpha^{4.9}_{1.4}$: it is 
consistent with peaking around observed 1.4~GHz (rest-frame 7~GHz) and 
due to the bandwidth narrowing with redshift, the frequencies observed 
range from rest-frame 25~GHz (observed 4.9~GHz) to rest-frame 3~GHz 
(observed 610~MHz).  In Fig.~\ref{fig:pz_plane} we can see that both 
AMS16 and AMS20 are intrinsically quite radio-faint at rest-frame 1.4 
GHz, placing an interesting limit on any associated 
massive-star-formation rate ($\lesssim$ 500-1000 M$_{\odot}$ 
yr$^{-1}$). 
 
AMS05 has $\alpha^{1.4}_{610} = 0.19$ and $\alpha^{4.9}_{1.4}=0.72$   
and lies in the `radio-intermediate' region of 
Fig.~\ref{fig:pz_plane}. The spectral indices can be explained by a 
GPS source peaking between 610 MHz  and  1.4 GHz (observed 
frequencies).

\section {Conclusions} 
 
We have used radio-data at three different frequencies, with 
well-matched beam sizes, to study the radio spectral properties of a 
population of high-redshift type-2 quasars.  The sample contains a 
range of different radio spectral properties, which include mainly 
steep-spectrum sources ($\alpha^{1.4}_{610}$ and 
$\alpha^{4.9}_{1.4}\sim1$), plus a minority of flat steep sources, 
with ($\alpha^{1.4}_{610} \lesssim 0.5$ or $\alpha^{4.9}_{1.4} 
\lesssim 0.5$), and three, or perhaps four, GHz-peaked 
sources. The presence of flat-spectrum sources would be hard to 
reconcile with obscuration by a torus but we find no secure examples 
of such a case as there are no narrow-line type-2s with  flat 
radio spectra. Otherwise, we find no correlation between radio 
spectral properties and optical spectral properties, suggesting no 
obvious dichotomy between ``host-obscured'' and ``torus-obscured'' 
type-2s, at least in terms of their radio spectral properties. 
 
When determining the radio luminosities of our sources, we find the 
majority fall below the FRI/FRII break but above the luminosity 
corresponding to a massive-star-formation rate of 1000 M$_{\odot}$ 
yr$^{-1}$. Although we do not generally have the surface brightness 
sensitivity to detect extended jets, many of these objects could have 
FRI-like structures.  Only one source has a radio luminosity 
consistent with being entirely due to a $\sim$ 300 M$_{\odot}$ 
yr$^{-1}$ starburst, consistent with its low-ionization-only emission 
lines. Only 4 out of 21 sources (at most 5 with AMS08) are 
flat-spectrum sources, so the idea that the radio-intermediate quasar 
population might be dominated by Doppler boosted radio-quiet quasars 
\citep[e.g][] {1993MNRAS.263..425M} is not correct for this high-$z$ 
population. 
 
Interestingly, we find that the steep spectrum ($\alpha\sim1$) sources 
have radio spectra consistent with those of low-redshift FRI 
sources once the effects of an increased CMB energy density is taken into account. The spectral indices can be explained by active developed jets with 
continuous injection of relativistic electrons and inverse-Compton 
scattering by CMB photons, leading to the `break frequency' being 
lowered to $\sim0.3$ GHz and hence $\alpha\sim1$ at observed frequencies above $\sim0.3/(1+z)$ GHz.

\section*{Acknowledgments} 
 
We thank Ian Heywood and Rob Ivison for communicating results prior to 
publication, Matt Jarvis for helping obtain the VLA observations 
and the staff of the GMRT. SR and TSG thank the UK PPARC for a Senior 
Research Fellowship and a Studentship respectively.  The VLA is a 
facility of the NRAO operated by Associated Universites, Inc., under 
co-operative agreement with the National Science Foundation. The GMRT 
is operated by the National Centre for Radio Astrophysics of the Tata 
Institute of Fundamental Research


\begin{thebibliography}{} 
 
\bibitem[\protect\citeauthoryear{{Antonucci}}{{Antonucci}}{1993}]{1993ARA&A..3%
1..473A} 
{Antonucci} R.,  1993, \araa, 31, 473 
 
\bibitem[\protect\citeauthoryear{{Blundell} \& {Rawlings}}{{Blundell} \& 
  {Rawlings}}{2001}]{2001ApJ...562L...5B} 
{Blundell} K.~M.,  {Rawlings} S.,  2001, \apjl, 562, L5 
 
\bibitem[\protect\citeauthoryear{{Brandt} \& {Hasinger}}{{Brandt} \& 
  {Hasinger}}{2005}]{2005ARA&A..43..827B} 
{Brandt} W.~N.,  {Hasinger} G.,  2005, \araa, 43, 827 
 
\bibitem[\protect\citeauthoryear{{Condon}}{{Condon}}{1992}]{1992ARA&A..30..575%
C} 
{Condon} J.~J.,  1992, \araa, 30, 575 
 
\bibitem[\protect\citeauthoryear{{Condon}, {Cotton}, {Yin}, {Shupe}, 
  {Storrie-Lombardi}, {Helou}, {Soifer} \& {Werner}}{{Condon} 
  et~al.}{2003}]{2003AJ....125.2411C} 
{Condon} J.~J.,  {Cotton} W.~D.,  {Yin} Q.~F.,  {Shupe} D.~L., 
  {Storrie-Lombardi} L.~J.,  {Helou} G.,  {Soifer} B.~T.,    {Werner} M.~W., 
  2003, \aj, 125, 2411 
 
\bibitem[\protect\citeauthoryear{{Fanaroff} \& {Riley}}{{Fanaroff} \& 
  {Riley}}{1974}]{1974MNRAS.167P..31F} 
{Fanaroff} B.~L.,  {Riley} J.~M.,  1974, \mnras, 167, 31P 
 
\bibitem[\protect\citeauthoryear{{Haas}, {Siebenmorgen}, {Schulz}, {Kr{\"u}gel} 
  \& {Chini}}{{Haas} et~al.}{2005}]{2005A&A...442L..39H} 
{Haas} M.,  {Siebenmorgen} R.,  {Schulz} B.,  {Kr{\"u}gel} E.,    {Chini} R., 
  2005, \aap, 442, L39 
 
\bibitem[\protect\citeauthoryear{{Kardashev}}{{Kardashev}}{1962}]{1962SvA.....%
6..317K} 
{Kardashev} N.~S.,  1962, Soviet Astronomy, 6, 317 
 
\bibitem[\protect\citeauthoryear{{Lacy} et~al.,}{{Lacy} 
  et~al.}{2004}]{2004ApJS..154..166L} 
{Lacy} M.,  et~al., 2004, \apjs, 154, 166 
 
\bibitem[\protect\citeauthoryear{{Laing} \& {Peacock}}{{Laing} \& 
  {Peacock}}{1980}]{1980MNRAS.190..903L} 
{Laing} R.~A.,  {Peacock} J.~A.,  1980, \mnras, 190, 903 
 
\bibitem[\protect\citeauthoryear{{Leahy}}{{Leahy}}{1991}]{1991bja..book..100L} 
{Leahy} J.~P.,  1991, Beams and Jets in Astrophysics, 100 
 
\bibitem[\protect\citeauthoryear{{Leipski}, {et al.}}{{Leipski} et~al.}{2005}]{2005A&A...440L...5L} 
{Leipski} C.,  et al.,  2005, \aap, 440, L5 
 
\bibitem[\protect\citeauthoryear{{Mart{\'{\i}}nez-Sansigre}, {Rawlings}, 
  {Lacy}, {Fadda}, {Jarvis}, {Marleau}, {Simpson} \& 
  {Willott}}{{Mart{\'{\i}}nez-Sansigre} et~al.}{2006}]{2006MNRAS.370.1479M} 
{Mart{\'{\i}}nez-Sansigre} A.,  {Rawlings} S.,  {Lacy} M.,  {Fadda} D., 
  {Jarvis} M.~J.,  {Marleau} F.~R.,  {Simpson} C.,    {Willott} C.~J.,  2006, 
  \mnras, 370, 1479 
 
\bibitem[\protect\citeauthoryear{{Mart{\'{\i}}nez-Sansigre}, {Rawlings}, 
  {Lacy}, {Fadda}, {Marleau}, {Simpson}, {Willott} \& 
  {Jarvis}}{{Mart{\'{\i}}nez-Sansigre} et~al.}{2005}]{2005Natur.436..666M} 
{Mart{\'{\i}}nez-Sansigre} A.,  {Rawlings} S.,  {Lacy} M.,  {Fadda} D., 
  {Marleau} F.~R.,  {Simpson} C.,  {Willott} C.~J.,    {Jarvis} M.~J.,  2005, 
  \nat, 436, 666 
 
\bibitem[\protect\citeauthoryear{{Miller}, {Rawlings} \& {Saunders}}{{Miller} 
  et~al.}{1993}]{1993MNRAS.263..425M} 
{Miller} P.,  {Rawlings} S.,    {Saunders} R.,  1993, \mnras, 263, 425 
 
\bibitem[\protect\citeauthoryear{{Morganti}, {Garrett}, {Chapman}, {Baan}, 
  {Helou} \& {Soifer}}{{Morganti} et~al.}{2004}]{2004A&A...424..371M} 
{Morganti} R.,  {Garrett} M.~A.,  {Chapman} S.,  {Baan} W.,  {Helou} G., 
  {Soifer} T.,  2004, \aap, 424, 371 
 
\bibitem[\protect\citeauthoryear{{O'Dea}}{{O'Dea}}{1998}]{1998PASP..110..493O} 
{O'Dea} C.~P.,  1998, \pasp, 110, 493 
 
\bibitem[\protect\citeauthoryear{Ogle, Whysong \& Antonucci}{Ogle 
  et~al.}{2006}]{Ogle:2006cu} 
Ogle P.~M.,  Whysong D.,    Antonucci R.,  2006 
 
\bibitem[\protect\citeauthoryear{{Rees}}{{Rees}}{1984}]{1984ARA&A..22..471R} 
{Rees} M.~J.,  1984, \araa, 22, 471 
 
\bibitem[\protect\citeauthoryear{{Weedman}, {Le Floc'h}, {Higdon}, {Higdon} \& 
  {Houck}}{{Weedman} et~al.}{2006}]{2006ApJ...638..613W} 
{Weedman} D.~W.,  {Le Floc'h} E.,  {Higdon} S.~J.~U.,  {Higdon} J.~L., 
  {Houck} J.~R.,  2006, \apj, 638, 613 
 
\bibitem[\protect\citeauthoryear{{Yan}, {et al.}}{{Yan} 
  et~al.}{2005}]{2005ApJ...628..604Y} 
{Yan} L.,  et al.,  2005, \apj, 628, 604 
 
\end{thebibliography}

\label{lastpage} 
 
\end{document}